\documentclass[reprint,aps,prd,floatfix,nofootinbib,showkeys,showpacs]{revtex4-2}
\usepackage{amsmath,amsthm,amssymb}
\usepackage{graphicx,color,dsfont,bm, url, hyperref}
\usepackage[utf8]{inputenc}
\def\W{\mathcal{W}}
\def\oW{\overline{\mathcal{W}}}
\def\cV{\mathcal{V}}
\def\inn{\mathrm{in}}
\def\out{\mathrm{out}}
\def\X{{\scriptscriptstyle\mathrm{X}}}
\def\R{{\scriptscriptstyle\mathrm{R}}}
\def\V{{\scriptscriptstyle\mathrm{V}}}

\def\s{\sigma}
\def\S{\Sigma}
\def\O{\Omega}
\def\Ob{\bar{\Omega}}

\renewcommand{\d}{\mathrm{d}}
\def\uh{\hat{u}}
\def\Uh{\hat{\mathcal{U}}}
\def\kt{\text{k}$_t$ }

\def\ktt{\scriptscriptstyle\text{k}_t}

\begin{document}

\title{Analytical structure of k$_t$ clustering to any order}

\author{K. \surname{Khelifa-Kerfa}}
\affiliation{Department of Physics, Faculty of Science and Technology\\
Relizane University, Relizane 48000, Algeria}
\email{kamel.khelifakerfa@univ-relizane.dz}

\begin{abstract}
We present a general analytical expression for the fixed-order structure of the distribution of a generic non-global observable with the \kt jet algorithm at any perturbative order. This novel formulation is obtained within the framework of the eikonal approximation, assuming strong-energy ordering of emitted partons. The proposed formula is applicable to a wide range of processes at both lepton and hadron colliders.
\end{abstract}


\maketitle

\section{Introduction}
\label{sec:Intro}

High-energy physics experiments, such as those conducted at the CERN Large Hadron Collider (LHC), often involve complex processes where particles are produced in large numbers and interact in a highly intricate manner. Among the essential tools used to understand the outcomes of these interactions are jet algorithms, which play a crucial role in identifying sets of particles that originate from the same high-energy (boosted) particle. The identification and reconstruction of jets is a fundamental aspect of collider experiments.

The longitudinally invariant \kt jet algorithm, first introduced in Refs. \cite{Catani:1993hr, Ellis:1993tq}, is one of the infrared and collinear (IRC) safe sequential recombination algorithms \cite{Cacciari:2011ma}, which play a pivotal role in jet reconstruction. Unlike other jet algorithms--such as the anti-\kt algorithm \cite{Cacciari:2008gp} and various non-iterative cone algorithms--that initiate jet formation from hard partons, the \kt algorithm prioritizes the clustering of soft partons first, i.e., those with the lowest relative transverse momentum. Given that many QCD observables, or at least a significant subset of them, are sensitive to the energy scales and parton types involved in the interaction, and are frequently defined based on the final-state jets, they are significantly influenced by the clustering procedure--particularly at small scales where soft gluon emissions dominate.

A special class of QCD observables that are directly and significantly affected by the application of jet algorithms are known as non-global observables. These observables are defined over restricted regions of the final-state phase space. In 2001, Dasgupta and Salam \cite{Dasgupta:2001sh, Dasgupta:2002bw} demonstrated for the first time that this exclusivity in phase space leads to the emergence of an infinite tower of large logarithms, termed non-global logarithms (NGLs), which had previously been overlooked for years. These logarithms were subsequently resummed to all orders numerically, in the large-N$_c$ limit (where N$_c$ is the number of colors), using Monte Carlo (MC) methods for various phase-space geometries \cite{Dasgupta:2001sh, Dasgupta:2002bw}, with no clustering imposed.

One year after the discovery of NGLs, Appleby and Seymour showed in 2002 \cite{Appleby:2002ke} that NGLs are significantly reduced, though not completely eliminated, when final-state partons are subject to the clustering condition of the invariant \kt algorithm. Moreover, it was later found in \cite{Banfi:2005gj} that while \kt clustering minimizes the impact of NGLs originating from correlated secondary gluon emissions, it introduces a new tower of large logarithms for primary gluon emissions, known as clustering logarithms (CLs) \footnote{These may also be referred to as Abelian NGLs \cite{Kelley:2012kj}.}. The latter were absent in the earlier work of Appleby and Seymour \cite{Appleby:2002ke, Appleby:2003sj}. CLs were computed analytically up to two-loops in \cite{Banfi:2005gj} and extended to four-loops in \cite{Delenda:2006nf} for the interjet-energy-flow (or gaps-between-jets) observable. Subsequently, CLs were computed at various orders in the perturbative distribution of other non-global observables (see, for instance, \cite{Banfi:2010pa, Khelifa-Kerfa:2011quw, Kelley:2012kj, Kelley:2012zs, Delenda:2012mm, Kerfa:2012yae, Benslama:2023gys}).

In all of the aforementioned references that studied \kt clustering, the all-orders resummation of CLs and NGLs were performed only numerically, and in the case of NGLs only in the large-N$_c$ limit, with the exception of Refs. \cite{Delenda:2006nf, Delenda:2012mm}, which presented a partial analytical resummation of CLs for the specific case of interjet-energy-flow and invariant jet mass in $e^+ e^-$ annihilation processes, respectively. It is worth noting that the MC program of \cite{Dasgupta:2001sh} includes \kt clustering for both primary and secondary emissions, but only for a limited set of observables. Unlike NGLs in the anti-\kt jet algorithm, which have been successfully resummed numerically accounting for the full color structure \cite{Weigert:2003mm, Hatta:2013iba, Hagiwara:2015bia, Hatta:2020wre, Platzer:2013fha, AngelesMartinez:2018cfz, Forshaw:2019ver}, NGLs with \kt clustering have resisted any improvements beyond the large-N$_c$ approximation. Recently, CLs and NGLs have also been addressed within the Soft and Collinear Effective Theory (SCET) framework \cite{Becher:2023znt} (for the specific interjet-energy-flow observable), where the authors identified two main effects of clustering: an increase in the jet size due to clustering of soft radiation into the jet, and a reduction of secondary radiation effects due to the clustering of small-angle (collinear) radiation.

In addition to being studied at lepton-lepton colliders, CLs have also been addressed for lepton-hadron \cite{Dasgupta:2002dc} and hadron-hadron colliders \cite{Ziani:2021dxr, Bouaziz:2022tik}. Moreover, unlike CLs, which have been computed analytically up to four-loops as mentioned earlier, the fixed-order analytical calculation of NGLs in the presence of clustering has only recently been extended to three loops for the special case of the azimuthal decorrelation observable at lepton colliders \cite{Benslama:2023gys}, after being stuck at two loops for quite some time.

In the present paper, we provide a general formula for the effect of \kt clustering at each order, and to any order, of the perturbative distribution of a generic non-global observable. Our work is based on eikonal (soft) theory and implements the strong-energy ordering condition on the transverse momenta (or energies) \footnote{We shall henceforth refer to the ordering as being in energy, but it should be understood that transverse momentum is equally meant, especially for hadronic collisions.} of the emissions: $Q \gg k_{t1} \gg k_{t2} \gg \cdots$, where $Q$ is the hard scale of the process. The latter significantly simplifies the calculations and is valid for single-logarithm accuracy (SLA). At this accuracy, we previously presented general formulas for the eikonal amplitudes squared for $e^+ e^-$ (two hard legs) \cite{Delenda:2015tbo} and vector/Higgs boson + jet (three hard legs) \cite{Khelifa-Kerfa:2020nlc} processes, with explicit expressions given for up to four-loops.

To construct the appropriate clustering constraints (or functions) at each order, we make use of the measurement operator technique, first introduced in \cite{Schwartz:2014wha} and subsequently extensively applied in \cite{Khelifa-Kerfa:2015mma}, together with the rigorous procedure used in \cite{Delenda:2006nf, Delenda:2012mm} (but in a more general way). The expressions derived herein apply to the full distribution of the non-global observable and contain, consequently, the appropriate clustering constraints that may be used to work out both CLs and NGLs contributions. Due to the symmetric pattern manifested at lower orders, the structural form of these expressions may, as we shall see in the main text, be straightforwardly and systematically generalized to any order in the perturbative (PT) expansion of the observable. Furthermore, they contain the full color (finite-N$_c$) dependence at each order and may be applied to both leptonic and hadronic processes. These two latter properties are directly inherited from the use of the eikonal amplitudes squared derived in \cite{Delenda:2015tbo, Khelifa-Kerfa:2020nlc}.

For hadronic processes involving two colored partons in the initial state, special attention must be given to the effect of Coulomb (or Glauber) gluons. As thoroughly explained in Refs.~\cite{Dokshitzer:2005ig, Forshaw:2006fk}, within the eikonal approximation, contributions to the (virtual) loop integral arise from two distinct regions:
The first, referred to as the ``eikonal gluon'' contribution, originates from the pole at \(k^2 = 0\). The residue of this pole is identical and opposite, in sign, to the phase space integral for the corresponding emission of a real soft gluon. A complete real-virtual cancellation occurs only outside the vetoed region \(\mathcal{V}\). Within \(\mathcal{V}\), a real-virtual mis-cancellation remains, leading to purely virtual contributions in the range \(Q^2 v < k_t^2 < Q^2\) (see Sec.~\ref{sec:Generalities} for details regarding \(\mathcal{V}\) and the veto \(v\)).
The second, denoted the ``Coulomb or Glauber gluon'' contribution, arises from a pole where one of the emitting partons is on-shell. This corresponds to a space-like gluon carrying only transverse momentum. The residue is purely imaginary and non-zero only if both emitting partons are either in the initial or final state, leading to the well-known \(\imath \pi\) terms in the scattering amplitude.

Ref.~\cite{Forshaw:2006fk} demonstrated that Glauber gluons induce super-leading logarithms (SLLs) exclusively in \(2 \to 2\) scatterings. However, more recent studies \cite{Becher:2021zkk, Becher:2023mtx, Boer:2024hzh} have revealed that SLLs appear in any \(2 \to l\) process (with \(l = 0, 1, 2\)), albeit with \(2 \to 0\) and \(2 \to 1\) contributions being numerically suppressed.

In the context of our work on \kt clustering effects, we emphasize the following:
First, Glauber gluons are virtual and are therefore unaffected by clustering algorithms. As demonstrated in the main text, clustering pertains exclusively to real emissions. The drag-in and drag-out effects involve only real gluons. Virtual gluons cannot drag, nor can they be dragged, by either real or virtual gluons. Consequently, the \(k_t\) clustering algorithm has no influence on contributions arising from Glauber gluons.
Second, SLLs are present for any non-global observable, irrespective of whether clustering is applied. Unlike NGLs, which are reduced by clustering, SLLs appear to be independent of clustering effects, as discussed above.
Nonetheless, a definitive confirmation of these observations, as well as a detailed analysis of potential interferences among various large logarithms, requires further investigation. These aspects will be addressed in our upcoming publications.

Although the current work focuses on the \kt clustering algorithm, the essential ideas and procedures are readily applicable to various other jet clustering algorithms. In particular, we reserve the derivation of the analogous formulas for the Cambridge/Aachen \cite{Dokshitzer:1997in, Wobisch:1998wt} and other cone-like jet algorithms for a future publication \cite{CAclus}.

This paper is organized as follows. In Sec. \ref{sec:Generalities}, we define the necessary ingredients that will be used in later sections. These include the non-global observable, the measurement operator, and the \kt jet algorithm. We then explicitly derive the observable distribution, or rather the integrand, as we do not carry out the integrations explicitly—this would otherwise require a specific observable definition—at two-, three-, and four-loops in Secs. \ref{sec:2loop}, \ref{sec:3loop}, and \ref{sec:4loop}, respectively. Based on the symmetry pattern observed at these loop orders, we present in Sec. \ref{sec:Conclusion} the corresponding general formula for any given loop order $n$. Finally, we draw our conclusions in Sec. \ref{sec:Conclusion}.

\section{Generalities}
\label{sec:Generalities}
\subsection{Observable and measurement operator}
\label{subsec:Definitions}

Consider a non-global observable, such as jet shapes, hemisphere mass, and gaps-between-jets, to mention a few, which is defined in terms of some function $V(k_1, \dots, k_n)$ of the final-state four-momenta $k_1, \dots, k_n$ for a given number $n$ of final-state soft, energy-ordered gluons.\footnote{The function $V$ may also depend on the four-momenta of the final-state hard partons, other than gluons, in which case such dependence is implicit in our notation.} We wish to compute the observable integrated cross-section, $\S(v)$, for which the \kt clustering is applied to the final-state partons, and the observable is vetoed to be less than some value $v$ in a specific region of phase space $\cV$:

\begin{multline}\label{eq:IntegratedXsec}
 \S^{\ktt}(v) = \int \frac{1}{\s_0} \frac{\d\s}{\d v'}\,\Theta\left[v - V(k_1, \dots, k_n) \right]\times
 \\
 \times  \Xi^{\ktt}(k_1, \dots, k_n) \, \d v',
\end{multline}
where $\s_0$ is the Born cross section and $\Xi^{\ktt}(k_q, \dots, k_n)$ is the resultant \kt clustering function that selects the phase space of the final-state parton configurations that contribute to the observable. The function $V(k_1, \dots, k_n)$ admits the following factorization, in the eikonal limit,
\begin{align}\label{eq:V- Factorization}
	V(k_1, \dots, k_n) = \sum_i V_i, \qquad V_i \equiv V(k_i).
\end{align}
The fixed-order PT expansion of $\S^{\ktt}(v)$ is given by:
\begin{align}\label{eq:IntegratedXsec-Expansion}
 \S^{\ktt}(v) = 1 + \S^{\ktt}_1(v) + \S^{\ktt}_2(v) + \cdots,
\end{align}
where the $m^\text{th}$ term in the expansion reads:
\begin{multline}\label{eq:IntegratedXsec-mOrder}
 \S^{\ktt}_m(v) = \sum_{\X} \int_{k_{t1} > k_{t2} > \cdots > k_{tm}}\, \left(\prod_{i=1}^m \d\Phi_i\right)\, \times
 \\
 \times \Uh_{m} \, \W_{12 \dots m}^{\X} \,\Xi^{\ktt}_m(k_1, \dots, k_m),
\end{multline}
where $\W^{\X}_{12 \dots m} = \W^{\X}(k_1, \dots, k_m)$ is the eikonal amplitude squared for the emission of $m$ energy-ordered, soft gluons in configuration $X$ normalized to the Born configuration as detailed in Refs. \cite{Delenda:2015tbo, Khelifa-Kerfa:2020nlc}, and $\d\Phi_i$ is the phase space factor for the $i^{\mathrm{th}}$ gluon. In Eq. \eqref{eq:IntegratedXsec-mOrder}, the sum is over all possible soft gluon configurations in which each gluon can either be real (R) or virtual (V). For instance, at one-loop, i.e., $m=1$ for the emission of only one gluon $k_1$, the possible gluon configurations are $X = \{\text{R, V}\}$. The sum in Eq. \eqref{eq:IntegratedXsec-mOrder} is then over the two eikonal amplitudes squared: $\W_1^\R$ (for which gluon $k_1$ is real) and $\W_1^\V$ (for which gluon $k_1$ is virtual).
 In the eikonal approximation and for energy-ordered emissions, the eikonal amplitude squared for gluon configurations in which the softest gluon is virtual is just the negative of that in which it is real \cite{Delenda:2015tbo, Khelifa-Kerfa:2020nlc}. That is:
\begin{align}\label{eq:RV-SoftGluonRelation}
 \W_{12 \dots m}^{\X_1 \X_2 \dots \V} = - \W_{12 \dots m}^{\X_1 \X_2 \dots \R},
\end{align}
where $\X_i$, corresponding to gluon $k_i$, may either be real R or virtual V. For the other features of the eikonal amplitudes squared the interested reader is referred to Refs. \cite{Delenda:2015tbo, Khelifa-Kerfa:2020nlc}.

The measurement operator $\Uh_m$, appearing in Eq. \eqref{eq:IntegratedXsec-mOrder}, is a non-linear operator acting on the eikonal amplitudes squared to ensure that only gluon configurations for which the observable $V(k_1, \dots, k_m)$, at $m^{\text{th}}$ order, is less than the value $v$ have a non-vanishing contribution to the integrated cross-section $\S_m^{\ktt}(v)$. All other gluon configurations contribute nothing. For strong-energy ordering, and after applying \kt clustering (i.e., once the final positions of the re-shuffled  partons are determined), its complicated form factorizes into a product of individual measurement operators for each soft gluon, i.e.,
\begin{align}
 \Uh_{m} = \prod_{i=1}^m \, \uh_i.
\end{align}
The eikonal amplitudes squared are then eigenfunctions of the operators $\uh_i$ with eigenvalues of 0 and 1. Eikonal amplitudes for configurations in which gluon $k_i$ is {\it real} and emitted {\it outside} the phase space region, $\cV$, in which the non-global observable is defined, are kept intact regardless of the nature of $k_i$. This is simply because being emitted outside $\cV$ means that $k_i$ does not contribute to the observable. If, however, $k_i$ is real and emitted {\it inside} the region $\cV$, then the eigenvalue of $\uh_i \W_{12 \dots m}^{\X}$ is 1 if $v > V_i$ and 0 otherwise. If gluon $k_i$ is virtual, then the eigenvalue is simply 1. Thus, the measurement operator $\uh_i$ for the $i^\text{th}$ emission may be written as follows \cite{Schwartz:2014wha, Khelifa-Kerfa:2015mma}:
\begin{align}\label{eq:uh_i}
 \uh_i = \Theta^\V_i + \Theta^\R_i \left[ \Theta^\out_i + \Theta^\inn_i \Theta(v - V_i) \right] = 1 - \Theta_i^v \Theta_i^\inn \Theta_i^\R,
\end{align}
where the various step functions are defined as follows:
\begin{itemize}
\item $\Theta_i^{\R(\V)}$ is 1 if gluon $k_i$ is real (virtual) and zero otherwise. We thus have $\Theta_i^\R + \Theta_i^\V = 1$.
\item  $\Theta_i^{\out(\inn)}$ is 1 if gluon $k_i$ is emitted outside (inside) the region $\cV$ and zero otherwise. We thus have $\Theta_i^\out + \Theta_i^\inn = 1 $.
\item $\Theta_i^v \equiv \Theta\left(V_i - v \right)$.
\end{itemize}
Using the above relations one may easily arrive at the second equality in Eq. \eqref{eq:uh_i}.

The details of how the \kt clustering function at $m^\text{th}$ order, $\Xi^{\ktt}_m$, is built up are discussed in the next section.

\subsection{The \kt jet algorithm}
\label{subsec:JetAlgo}

The (inclusive variant of the) longitudinally invariant \kt jet algorithm \cite{Catani:1993hr, Ellis:1993tq}, a specific case ($p=1$ below) of the generalized \kt algorithm \cite{Cacciari:2011ma}, may be defined as follows \cite{Cacciari:2011ma}:
\begin{enumerate}
\item For each pair of particles\footnote{Particles (or partons) are used in PT calculations. For experimental analyzes, particle tracks, cells and towers in actual detectors are used instead.} $i,j$, from the list of final-state particles, compute the distance
\begin{align}
 d_{ij} = \min\left(k_{ti}^{2p}, k_{tj}^{2p}\right) \, \Delta R_{ij}^2/R^2,
\end{align}
and for each particle $i$ compute the distance
\begin{align}
	d_{iB} = k_{ti}^{2p},
\end{align}
where $\Delta R_{ij}^2 = (\eta_i - \eta_j)^2 + (\phi_i - \phi_j)^2$ with $k_{ti}, \eta_i$ and $\phi_i$ being the transverse momentum, rapidity and azimuthal angle of particle $i$ with respect to the beam direction $B$. The variable $R$ is the jet-radius parameter.

\item If the minimum distance of all $d_{ij}$'s and $d_{iB}$'s is a $d_{ij}$, then the pair of particles $i,j$ is merged into a single particle (or pseudo-jet) with a four-momentum equal to the sum of the four-momenta of $i$ and $j$ (according to the E-scheme). In the strong-energy ordering limit, the pseudo-jet is essentially in the direction of the hardest particle.
If, however, the smallest distance of all is $d_{iB}$, then particle $i$ is declared a jet and is thus removed from the list of final-state particles.
\item Steps 1 and 2 are repeated until no particles are left in the initial list.
\end{enumerate}

For our analytical calculations, at any given PT order, we consider {\it all} clustering possibilities among the final-state partons, along with all possible initial gluon configurations (i.e., which ones are initially inside $\cV$ and which ones are initially  outside $\cV$). We then analyze each of these possibilities and identify those that would lead to mis-cancellations between real and virtual corrections, thus contributing to the non-global observable (and resulting in the appearance of CLs and/or NGLs). At three-loops and beyond, the set of all possibilities becomes lengthy and cumbersome, so we use \texttt{Mathematica} to handle the automated work.

At one-loop, i.e., the first and simplest case, \kt clustering has no impact on the observable distribution, which is thus identical to the anti-\kt clustering case. This is because for a single soft emission $k_1$ off the Born configuration, if $k_1$ is (real and) emitted inside the vetoed region $\cV$, then it contributes to the observable function $V(k_1)$; otherwise, it does not. In terms of the measurement operator, we write this as
\begin{align}\label{eq:1loop-uW1}
 \sum_\X \uh_1 \W_1^\X
 &= \uh_1 \W^\R_1 + \uh_1 \W^\V_1, \notag\\
 &= \W^\R_1 - \Theta^v_1 \Theta^\inn_1 \W^\R_1 + \W_1^\V =- \Theta^v_1 \Theta^\inn_1 \W^\R_1.
\end{align}
The last equality follows from the relation $\W^\R_1 + \W^\V_1 = 0$ (as in Eq. \eqref{eq:V- Factorization}). Notice that applying the \kt clustering does not modify this result as there are no other soft gluons to drag $k_1$ in or out of the vetoed region $\cV$. The effect of \kt  clustering begins at the two-loop order, which we address in the next section.

\section{Two-loops}
\label{sec:2loop}

Consider the emission of two soft energy-ordered gluons, $k_1$ and $k_2$ ($Q \gg k_{t1} \gg k_{t2}$), off a set of hard partons comprising the Born configuration. There are four distinct initial configurations for the aforementioned soft gluons. These are:
\begin{enumerate}
\item Both gluons, $k_1$ and $k_2$, are emitted inside the vetoed region $\cV$.
\item Both gluons, $k_1$ and $k_2$, are emitted outside the vetoed region $\cV$.
\item The harder gluon $k_1$ is emitted inside the vetoed region $\cV$, and the softer gluon $k_2$ is emitted outside the vetoed region $\cV$.
\item The harder gluon $k_1$ is emitted outside the vetoed region $\cV$, and the softer gluon $k_2$ is emitted inside the vetoed region $\cV$.
\end{enumerate}
If {\it no} \kt clustering is applied, then only the first configuration would contribute to the observable for primary emissions\footnote{These are gluons emitted directly off the Born hard partons (see Refs. \cite{Delenda:2015tbo, Khelifa-Kerfa:2020nlc} for more details).}, and only the last configuration would contribute to the observable for secondary correlated emissions\footnote{These are gluons emitted off other harder gluons.}. We first focus on the primary emissions case and then move on to the secondary correlated emissions case.

To clearly see why only the first configuration (case 1 above) contributes, let us apply the measurement operator to the corresponding eikonal amplitudes squared:
\begin{align}\label{eq:2loop-NoClus-1in2in}
 \sum_\X \Uh_2 \W^\X_{12} &= \uh_1 \uh_2 \left[\W^{\R\R}_{12} + \W^{\R\V}_{12} + \W^{\V\R}_{12} + \W^{\V\V}_{12}\right],
 \notag\\
 &= - \Theta_1^v \Theta_2^v\, \Theta_1^\inn \Theta_2^\inn \,\W^{\V\R}_{12}.
\end{align}
Notice that in passing from the first line to the second in the above equation, we made use of both eikonal approximation and strong-energy ordering. The eikonal approximation allows us to write $\W^{\R\V}_{12} = - \W^{\R\R}_{12}$ and $\W^{\V\V}_{12} = - \W^{\V\R}_{12}$ (in accordance with Eq. \eqref{eq:RV-SoftGluonRelation}), while strong-energy ordering allows us to write $\Theta(V_1 + V_2 - v) \approx \Theta(V_1 - v)$ and $\Theta_2^v = \Theta_2^v \Theta_1^v$. Physically, when both gluons are emitted inside the vetoed region $\cV$, both contribute to the observable with the constraint $\Theta\left(v - V_1 - V_2\right) \approx \Theta\left(v - V_1\right)$. If gluon $k_1$ is virtual, then only gluon $k_2$ contributes with a constraint $-\Theta(v - V_2)$, and vice versa with the constraint $-\Theta(v - V_1)$ (the minus sign is due to the fact that these are virtual corrections). If both gluons are virtual, then neither contributes. Adding up these contributions, one easily sees that the resultant constraint is $-\Theta(v - V_2)$, coming from the configuration where $k_1$ is virtual and $k_2$ is real. This is identically the expression arrived at in Eq. \eqref{eq:2loop-NoClus-1in2in} (after adding the last term $\W_{12}^{\V\V}$). This fact is evident in the explicit form of the eikonal amplitude squared appearing in Eq. \eqref{eq:2loop-NoClus-1in2in}: $\W_{12}^{\V\R} = \W_1^\R \W_2^\R$ (see Refs. \cite{Delenda:2015tbo, Khelifa-Kerfa:2020nlc}). In fact, the contribution from this configuration is nothing but the square of the one-loop contribution \eqref{eq:1loop-uW1} (as has been previously shown in, for example, \cite{Delenda:2006nf, Khelifa-Kerfa:2011quw, Delenda:2012mm}). By the same reasoning, we can easily convince ourselves that the other three gluon configurations have vanishing contributions to the observable distribution.

For the second case, i.e., secondary correlated emissions, the application of the measurement operator leads to:
\begin{align}\label{eq:2loop-NoClus-1out2in}
\sum_\X \Uh_2 \W^\X_{12} &= - \Theta_1^v \Theta_2^v\, \Theta_1^\out \Theta_2^\inn \left(\W_{12}^{\R\R} + \W_{12}^{\V\R}\right).
\end{align}
From Refs. \cite{Delenda:2015tbo, Khelifa-Kerfa:2020nlc}, we see that the sum $\W_{12}^{\R\R} + \W_{12}^{\R\V} = \oW_{12}^{\R\R}$ is proportional to the emission factor of two correlated gluons. Physically, gluon $k_1$, being harder and outside the vetoed region $\cV$, emits the softer gluon $k_2$ inside $\cV$, which thus contributes to the observable with the constraint $\Theta(v - V_2)$. The virtual correction to this contribution corresponds to gluon $k_2$ being virtual, which thus does not contribute to the observable. Hence, a mis-cancellation takes place, resulting in a contribution with the constraint $-\Theta(V_2 - v) = -\Theta_2^v$. This result is represented mathematically in Eq. \eqref{eq:2loop-NoClus-1out2in}. In fact, this very contribution is what led to the appearance of NGLs in the pioneering work of \cite{Dasgupta:2001sh, Dasgupta:2002bw}.

By the same reasoning (together with the strong-energy ordering condition), one can straightforwardly show that the other three initial configurations (1, 2, and 3 above) lead to no contributions to the observable distribution for the case of secondary correlated emissions. Therefore, in the case of no clustering (i.e., analogous to the case where anti-\kt or other cone-like jet algorithms are applied), the two-loop distribution of the non-global observable is proportional to (adding up Eqs. \eqref{eq:2loop-NoClus-1in2in} and \eqref{eq:2loop-NoClus-1out2in} and simplifying):
\begin{align}\label{eq:2loop-NoClus-Final}
 \sum_\X \Uh_2 \W_{12}^\X = -\Theta_1^v \Theta_2^v \Theta_2^\inn \left[\W_{12}^{\V\R} + \Theta_1^\out \W_{12}^{\R\R}  \right].
\end{align}
An identical expression has been derived in Ref. \cite{Khelifa-Kerfa:2015mma} for the particular case of the hemisphere mass observable.

If clustering is {\it turned on}, then upon applying the \kt jet algorithm to each of the above four initial configurations, we find for the {\it primary emissions} case:
\begin{enumerate}
	\item Clustering has no effect on the final result, since even if the two gluons are clustered together, they would still remain inside the vetoed region and thus both contribute to the observable. We thus obtain a contribution identical to that of Eq. \eqref{eq:2loop-NoClus-1in2in}.

	\item Clustering again has no effect, since even if the two gluons are clustered together, they would still remain outside the vetoed region and thus do not contribute to the observable.

	\item Although the harder gluon $k_1$ drags the softer gluon $k_2$ into the vetoed region (when both are real), the value of the observable does not change (after adding up the virtual corrections) due to the strong-energy ordering condition.

	\item When both gluons $k_1$ and $k_2$ are real, the former drags the latter out of the vetoed region if $d_{12} < d_{2B}$, and they are thus clustered together. This nullifies the contribution from this particular configuration, which would otherwise be present if there were no clustering. If, however, either of the two gluons (or both) is virtual, then no dragging is possible, and hence no clustering occurs. This situation creates a real-virtual mis-cancellation, resulting in a new contribution to the observable. Applying the measurement operator, we obtain:
\begin{align}\label{eq:2loop-ClusPrimary-1out2in}
\sum_\X \Uh_2 \W^\X_{12} &= - \Theta_1^v \Theta_2^v\, \Theta_1^\out \Theta_2^\inn\, \O_{12}\, \W_{12}^{\V\R},
\end{align}
where $\O_{ij} \equiv \Theta\left(d_{j} - d_{ij} \right)$ (with $d_j \equiv d_{jB}$) represents the clustering condition. We also represent the opposite (no clustering) condition as: $\Ob_{ij} = 1 - \O_{ij} \equiv \Theta\left(d_{ij} - d_{j}\right)$.
\end{enumerate}

For the correlated secondary emissions, only the last configuration (case 4 above) is modified by clustering; the other three remain unchanged. For this configuration, if both gluons are real and $d_{2} > d_{12}$, then $k_2$ is dragged by $k_1$ out of the vetoed region, leading to no contribution to the observable. The corresponding virtual correction also has a vanishing contribution. Adding up the two, we find that clustering has nullified any contribution from this particular initial configuration. Consequently, if this gluon configuration is to contribute to the value of the observable, it must {\it survive} the clustering; that is, gluon $k_2$ must be closer to the beam than to gluon $k_1$: $d_{2} < d_{12}$. Applying the measurement operator, we find:
\begin{align}\label{eq:2loop-Clus-1out2in}
\sum_\X \Uh_2 \W^\X_{12} &= - \Theta_1^v \Theta_2^v\, \Theta_1^\out \Theta_2^\inn\, \Ob_{12} \left(\W_{12}^{\R\R} + \W_{12}^{\V\R}\right).
\end{align}

The two possibilities of clustering discussed above, i.e., $d_{12} < d_2$ and $d_{12} > d_{2}$ (or equivalently $\O_{12} = 1 \; (\Ob_{12} = 0)$ and $\O_{12} = 0\; (\Ob_{12} = 1)$), are represented in Table \ref{tab:2loop}.
\begin{table}[t]
\centering
\caption{Clustering possibilities at two-loops with various initial gluon configurations. The letters "C" and ``c" refer to real-virtual mis-cancellations for primary and secondary emissions, respectively, and thus a contribution to the value of the observable.} \label{tab:2loop}
\begin{tabular}{c||c|c|c|c}
\hline
$\O_{12}$ & $\Theta_1^\inn \Theta_2^\inn$ & $\Theta_1^\out \Theta_2^\out$ & $\Theta_1^\inn \Theta_2^\out$ & $\Theta_1^\out \Theta_2^\inn$  \\
\hline \hline
$0$ & C & $0$ & $0$ & c \\
\hline
$1$ & C & $0$ & $0$ & C \\
\hline
\end{tabular}
\end{table}

Summing up all contributions when clustering is demanded (which corresponds to summing all rows in Table \ref{tab:2loop}) and simplifying we obtain:
\begin{multline}\label{eq:2loop-Clus-Final}
\sum_\X \Uh_2\, \W_{12}^\X = -\Theta_1^v \Theta_2^v \Theta_2^\inn
\left[ \W_{12}^{\V\R} + \Theta_1^\out\,\Ob_{12}\, \W_{12}^{\R\R}  \right].
\end{multline}
A few important points to spell out regarding the above formula:
\begin{itemize}
	\item Gluon configurations that have non-vanishing contributions to the observable distribution must always have the {\it softest} gluon emitted {\it inside} the vetoed region (i.e., $k_2$ in the current two-loop case).

	\item Gluon configurations where all gluons are emitted inside the vetoed region always contribute to the observable distribution, regardless of clustering. Conversely, gluon configurations where all gluons are emitted outside the vetoed region never contribute to the observable distribution, regardless of clustering.

	\item In the case of no clustering, that is, $\O_{12} = 0$ (and hence $\Ob_{12} = 1$), Eq. \eqref{eq:2loop-Clus-Final} reduces to Eq. \eqref{eq:2loop-NoClus-Final}, as expected.

	\item If we focus on the harder gluon $k_1$ (and ignore the softest gluon $k_2$, for the moment, which is always real and inside the vetoed region), we see from Eq. \eqref{eq:2loop-Clus-Final} that there are two terms corresponding to $k_1$ being virtual (the first term) and real (the second term). Whenever the harder gluon is virtual, the notions of inside and outside, and clustered and not clustered, do not apply. This is why no factor is multiplying the corresponding eikonal amplitude squared $\W_{12}^{\V\R}$. Conversely, if $k_1$ is real, it should always be outside the vetoed region $\cV$ and survive the clustering with the softest gluon. This is clearly seen in the factor multiplying $\W_{12}^{\R\R}$.

	\item Substituting the eikonal amplitudes squared from Refs. \cite{Delenda:2015tbo, Khelifa-Kerfa:2011quw}, we can easily show that the resultant expression for primary emissions coincides with that reported in Refs. \cite{Delenda:2006nf, Delenda:2012mm}.

	\item From Eqs. \eqref{eq:2loop-NoClus-1in2in}, \eqref{eq:2loop-ClusPrimary-1out2in}, and \eqref{eq:2loop-Clus-1out2in} (or alternatively from Eq. \eqref{eq:2loop-Clus-Final} using the relations $\Theta_i^\inn + \Theta_i^\out = 1$ and $\O_{ij} + \Ob_{ij} = 1$), we can extract the clustering functions for both primary and secondary correlated emissions that are responsible for CLs and NGLs, respectively. They read:
\begin{subequations}
\begin{flalign}
 \Xi_{\text{\tiny CLs}}^{\ktt}(k_1, k_2) &= \Theta_1^\out \Theta_2^\inn\,\O_{12}, \\
 \Xi_{\text{\tiny NGLs}}^{\ktt}(k_1, k_2) &= \Theta_1^\out \Theta_2^\inn\,\Ob_{12}.
\end{flalign}
\end{subequations}
These are again identical to the findings of previous works, such as \cite{Delenda:2006nf, Khelifa-Kerfa:2011quw, Delenda:2012mm}.
\end{itemize}
In the next section we extend the above calculations to the three-loops order.

\section{Three-loops}
\label{sec:3loop}

For the emission of three energy-ordered soft gluons, $k_1$, $k_2$, and $k_3$, we divide them into three distinct pairs: $(12), (13)$ and $(23)$. Since each pair may either undergo ($\O_{ij} = 1$) or survive ($\O_{ij} = 0$) the \kt clustering, we have a total of eight possibilities. There are only three initial gluon configurations that need to be considered, based on the findings at two-loops. These, along with the former possibilities, are displayed in Table \ref{tab:3loop}. The other possible configurations have trivial results. In particular, if all three gluons are emitted inside the vetoed region, then such a configuration would contribute to the observable value regardless of clustering. Conversely, if all three gluons are outside the vetoed region, this configuration does not contribute, regardless of clustering.
\begin{table}[t]
\centering
\caption{Clustering possibilities at three-loops with various initial gluon configurations. For initial gluon configurations we only show gluons that are inside the vetoed region. Any gluon not explicitly shown means that it is outside the said region. The notation ``Cc" refers to real-virtual mis-cancellations for both primary and secondary emissions.} \label{tab:3loop}
\begin{tabular}{c|c|c||c|c|c}
\hline
$\O_{23}$ & $\O_{13}$ & $\O_{12}$ & $\Theta_3^\inn$ & $\Theta_2^\inn \Theta_3^\inn$ & $\Theta_1^\inn \Theta_3^\inn$ \\
\hline \hline
$0$ & $0$ & $0$ & c & c  & c \\
\hline
$0$ & $0$ & $1$ & c & Cc & c \\
\hline
$0$ & $1$ & $0$ & c & Cc & c \\
\hline
$0$ & $1$ & $1$ & c & Cc & c \\
\hline
$1$ & $0$ & $0$ & c & c  & Cc \\
\hline
$1$ & $0$ & $1$ & c & c  & Cc \\
\hline
$1$ & $1$ & $0$ & Cc & Cc & Cc \\
\hline
$1$ & $1$ & $1$ & Cc & Cc & Cc \\
\hline
\end{tabular}
\end{table}

Applying the measurement operator on the eikonal amplitudes squared corresponding to each of the possible configurations in Table \ref{tab:3loop} we find:
\begin{align}\label{eq:3loop-uWX3}
\sum_X \Uh_3 \W_{123}^\X &= -\Theta_1^v \Theta_2^v \Theta_3^v \Theta_3^\inn \Big[
 \Theta_1^\inn \Theta_2^\inn\, \W_{123}^{\V\V\R} +\notag\\
&+\Theta_1^\out \Theta_2^\out \big( \Ob_{13} \Ob_{23} \W_{123}^{\R\R\R}  +\Ob_{13} \W_{123}^{\R\V\R} \notag\\&+ \Ob_{23} \W_{123}^{\V\R\R} + \W_{123}^{\V\V\R}  \big)
\notag\\
&+\Theta_1^\out \Theta_2^\inn \big(\O_{12} \Ob_{13} \Ob_{23} \W_{123}^{\R\R\R} + \Ob_{13} \W_{123}^{\R\V\R} \notag\\&+ \W_{123}^{\V\V\R}  \big)
\notag\\
&+ \Theta_1^\inn \Theta_2^\out \left(\Ob_{23} \W_{123}^{\V\R\R} + \W_{123}^{\V\V\R}  \right).
\end{align}
This expression can be simplified and recast in the following form:
\begin{align}\label{eq:3loop-Clus-Final}
\sum_\X \Uh_3\, \W_{123}^\X &= -\Theta_1^v \Theta_2^v \Theta_3^v \Theta_3^\inn \Big[ \W_{123}^{\V\V\R} + \Theta_1^\out\,\Ob_{13}\, \W_{123}^{\R\V\R} + \notag\\
&+\Theta_2^\out\,\Ob_{23}\, \W_{123}^{\V\R\R} + \notag\\
&+ \Theta_1^\out \left(\Theta_2^\out + \Theta_2^\inn\,\Omega_{12}\right) \Ob_{13} \Ob_{23} \W_{123}^{\R\R\R}
\Big].
\end{align}
Comparing this to the two-loop result \eqref{eq:2loop-Clus-Final}, we can easily confirm all of the properties highlighted for the latter two-loop case. In particular, the terms in the expression \eqref{eq:3loop-Clus-Final} correspond to the permutations of $\{\text{R,V} \}$ for the two harder gluons $k_1$ and $k_2$. In other words, these two gluons can be in one of the following paired configurations: VV, RV, VR, and RR. Moreover, whenever any of them is real, it should be outside the vetoed region $\cV$ and survive the clustering with the softest gluon $k_3$. When both of them are real (the last term in Eq. \eqref{eq:3loop-Clus-Final}), we see that the softer among them, i.e., gluon $k_2$, can be outside $\cV$ in two different ways: \begin{itemize}
	\item Initially being outside $\cV$ and staying there after applying the clustering. This is represented by the first term, $\Theta_2^\out$, inside the bracket of the last line in Eq. \eqref{eq:3loop-Clus-Final}.
	\item Initially being inside $\cV$ and then dragged outside by the slightly harder gluon $k_1$ after applying the clustering. This is represented by the second term, $\Theta_2^\inn\, \O_{12}$, inside the bracket of the last line in Eq. \eqref{eq:3loop-Clus-Final}.
\end{itemize}
If we swap the energy ordering between the two gluons, such that $k_2$ is slightly harder than $k_1$, then we would have, in the last line of Eq. \eqref{eq:3loop-Clus-Final}, the following factor instead: $\Theta_2^\out \left(\Theta_1^\out + \Theta_1^\inn \O_{12} \right)$.

It is plausible that in the case of no clustering, i.e., $\O_{12} = \O_{13} = \O_{23} = 0$ (or equivalently $\Ob_{12} = \Ob_{13} = \Ob_{23} = 1$), Eq. \eqref{eq:3loop-Clus-Final} reduces to the corresponding expression in the anti-\kt (no clustering) case, as reported in Eq. (3.2) of \cite{Khelifa-Kerfa:2015mma}. Moreover, substituting the eikonal amplitudes squared, one can verify that the clustering functions for primary and secondary correlated emissions\footnote{Their explicit forms for the dijet mass observable are to be found in our paper \cite{Khelifa-Kerfa:2024hwx}.} are identical to those found in \cite{Delenda:2006nf, Delenda:2012mm, Benslama:2023gys} and \cite{Benslama:2023gys}, respectively, which were derived for specific observables. In fact, Ref. \cite{Benslama:2023gys} is the only work in the literature, that we are aware of, that has successfully computed the distribution of a non-global observable (specifically the dijet azimuthal decorrelation in $e^+ e^-$ collisions) at fixed order up to three-loops with \kt clustering. Our formula \eqref{eq:3loop-Clus-Final} is general and comprehensive, in the sense that it may be utilized for any non-global observable and contains all necessary terms responsible for both CLs and NGLs.

Beyond this loop order, only contributions from primary emissions are known (up to four-loops only) with \kt clustering. No similar information is known for secondary emission contributions. In the next section, we derive the formula for the observable distribution with \kt clustering at four-loops, which allows for such contributions to be computed for any non-global observable.

\section{Four-loops}
\label{sec:4loop}

From the symmetric pattern observed at the two- and three-loop orders (particularly in Eqs. \eqref{eq:2loop-Clus-Final} and \eqref{eq:3loop-Clus-Final}), it is possible to infer the full formula for the observable distribution at four-loops without any further work. Nonetheless, we have explicitly carried out the calculations to verify the inferred expression. To this end, we have six possible paired configurations for the four soft energy-ordered gluons $k_1, k_2, k_3$, and $k_4$: $\{(12), (13), (14), (23), (24), (34)\}$. Each pair can either undergo or survive the clustering, resulting in a total of $64$ distinct cases. The relevant possible initial gluon configurations at this loop order are: $\Theta_4^\inn$, $\Theta_4^\inn \Theta_3^\inn$, $\Theta_4^\inn \Theta_2^\inn$, $\Theta_4^\inn \Theta_1^\inn$, $\Theta_4^\inn \Theta_3^\inn \Theta_2^\inn$, $\Theta_4^\inn \Theta_3^\inn \Theta_1^\inn$, and $\Theta_4^\inn \Theta_2^\inn \Theta_1^\inn$ (gluons not explicitly stated in each case are assumed to be outside. For instance, the first case $\Theta_4^\inn$ is equivalent to $\Theta_4^\inn \Theta_3^\out \Theta_2^\out \Theta_1^\out$).

The application of the measurement operator on the corresponding eikonal amplitudes squared simplifies to:
\begin{align}\label{eq:4loop-Clus-Final}
\sum_\X \Uh_4 \W_{1234}^\X &=- \Theta_1^v \Theta_2^v \Theta_3^v \Theta_4^v \Theta_4^\inn \Big[
\W_{1234}^{\V\V\V\R} \notag\\&
+ \Theta_1^\out\,\Ob_{14}\, \W_{1234}^{\R\V\V\R}
\notag\\&
+ \Theta_2^\out\,\Ob_{24}\,\W_{1234}^{\V\R\V\R}
+ \Theta_3^\out\,\Ob_{34}\, \W_{1234}^{\V\V\R\R}
\notag\\&
+ \Theta_1^\out \left(\Theta_2^\out + \Theta_2^\inn \O_{12} \right)\Ob_{14} \Ob_{24}\, \W_{1234}^{\R\R\V\R}
\notag\\&
+ \Theta_1^\out \left(\Theta_3^\out + \Theta_3^\inn \O_{13} \right)\Ob_{14} \Ob_{34}\, \W_{1234}^{\R\V\R\R}
\notag \\&
+ \Theta_2^\out \left(\Theta_3^\out + \Theta_3^\inn \O_{23} \right)\Ob_{24} \Ob_{34}\, \W_{1234}^{\V\R\R\R}
\notag\\
&+ \Theta_1^\out \left(\Theta_2^\out + \Theta_2^\inn \O_{12} \right) \times
\notag\\&
\left(\Theta_3^\out + \Theta_3^\inn \left[\O_{23} + \Ob_{23} \O_{13} \right] \right) \times
\notag\\&
\times \Ob_{14} \Ob_{24} \Ob_{34}\, \W_{1234}^{\R\R\R\R}
\Big].
\end{align}
As anticipated, the various terms in the above expression correspond to the possible real/virtual configurations of the three harder gluons $k_1$, $k_2$, and $k_3$ (ignoring the softest gluon $k_4$, which is always real and inside $\cV$). That is: VVR, RVV, VRV, VVR, RRV, RVR, VRR, and RRR. The last term in Eq. \eqref{eq:4loop-Clus-Final}, where all three gluons are real and hence should all be outside the vetoed region $\cV$, can be understood as follows:
\begin{itemize}
	\item The hardest gluon, $k_1$, is initially outside $\cV$ and remains so after clustering, as it cannot be dragged by other softer gluons.
	\item The next-to-hardest gluon, $k_2$, can either be initially outside or inside $\cV$. After clustering, it will end up outside $\cV$ in both cases: it stays outside in the first case (corresponding to the factor $\Theta_2^\out$) and is dragged out by $k_1$ in the second case (corresponding to the factor $\Theta_2^\inn \O_{12}$).
	\item The next-to-next-to-hardest gluon, $k_3$, can end up outside $\cV$ in three different ways: initially outside $\cV$ and staying outside since the harder gluons are outside; initially inside $\cV$ and being dragged out after clustering by either $k_2$ or $k_1$. Since \kt clustering deals with softer gluons first, $k_3$ is dragged by $k_2$ whenever possible before it can be dragged by $k_1$. This is why we have the factor $\O_{13} \Ob_{23}$ in the penultimate line of \eqref{eq:4loop-Clus-Final}, which means that $k_3$ must survive clustering with $k_2$ before it can be clustered with $k_1$. This can be made more transparent by expanding the factor multiplying $\Ob_{14} \Ob_{24} \Ob_{34} \W_{1234}^{\R\R\R\R}$ and closely examining each term.
\end{itemize}

In the no-clustering case ($\O_{ij} = 0$, or equivalently $\Ob_{ij} = 1$ for all pairs $ij$ with $i,j = 1,2,3,4$), one recovers the corresponding expression in the anti-\kt (no clustering) case, as written in Eq. (3.15) of \cite{Khelifa-Kerfa:2015mma}. For primary emissions, and upon substituting the various eikonal amplitudes squared, one can readily show that the resultant expression is identical to that found in Refs. \cite{Delenda:2006nf, Delenda:2012mm}.

It is worth mentioning that the computation of CLs, which originate from clustering effects on primary emissions, has not been performed beyond this loop order (NGLs calculations have stopped at three-loops). Therefore, the results presented in the next section, which extend beyond four-loops, have, to the best of our knowledge, not been considered in the literature before, neither for CLs nor (of course) for NGLs, and are thus completely new.

\section{Beyond four-loops}
\label{sec:nloop}

Based on the patterns observed at previous loop orders, we can derive a general formula for the distribution of a non-global observable at any loop order $n$ when \kt clustering is applied to the final-state partons, utilizing measurement operator techniques. The formula is as follows:
\begin{widetext}
\begin{align}\label{eq:nloop-Clus-Final}
\sum_\X \Uh_n \W_{1 \dots n}^{\X} &= -\prod_{i=1}^{n} \Theta_i^v \, \Theta_n^\inn \Bigg[
 \W_{1 \dots n}^{\{\R_n\}}
+ \sum_{j=1}^{n-1} \Theta_j^\out \,\Ob_{jn}\, \W_{1 \dots n}^{ \{\R_j,\R_n\}}
\notag\\&
+ \sum_{1\leq j<k}^{n-1} \Theta_{j}^\out \left(\Theta_{k}^\out + \Theta_{k}^\inn \O_{j k} \right) \Ob_{j n} \Ob_{k n} \,\W_{1 \dots n}^{\{\R_{j},\R_{k},\R_n\}}
\notag\\&
+ \sum_{1\leq j<k<\ell}^{n-1} \Theta_j^\out \left(\Theta_k^\out + \Theta_k^\inn \O_{jk} \right) \left(\Theta_\ell^\out + \Theta_\ell^\inn \left[\O_{k\ell} + \Ob_{k \ell} \O_{j\ell} \right] \right) \Ob_{jn} \Ob_{kn} \Ob_{\ell n}\, \W_{1 \dots n}^{\{\R_j,\R_k,\R_\ell,\R_n\}}
\notag\\&
+  \sum_{1\leq j<k<\ell<m}^{n-1} \Theta_j^\out \left(\Theta_k^\out + \Theta_k^\inn \O_{jk} \right) \left(\Theta_\ell^\out + \Theta_\ell^\inn \left[\O_{k\ell} + \Ob_{k \ell} \O_{j\ell} \right] \right) \notag\\&\times \left(\Theta_m^\out + \Theta_m^\inn \left[\O_{\ell m} + \Ob_{\ell m} \O_{k m} + \Ob_{\ell m} \Ob_{k m} \O_{j m} \right] \right) \Ob_{jn} \Ob_{kn} \Ob_{\ell n} \Ob_{mn}\, \W_{1 \dots n}^{\{\R_j,\R_k,\R_\ell,\R_m,\R_n\}}
\notag\\&
+ \cdots
\notag\\&
+ \Theta_{1}^\out  \left(\Theta_{2}^\out + \Theta_{2}^\inn \O_{12} \right) \cdots \big(\Theta_{n-1}^\out + \Theta_{n-1}^\inn \big[ \O_{(n-2)(n-1)} + \Ob_{(n-2)(n-1)} \O_{(n-3) (n-1)} + \cdots \notag\\&
\dots + \Ob_{(n-2)(n-1)} \Ob_{(n-3)(n-1)} \dots \Ob_{2(n-1)} \O_{1(n-1)}  \big] \big) \Ob_{1n} \Ob_{2n} \cdots \Ob_{(n-1)n}\, \W_{1 \dots n}^{\R_{1} \R_{2} \dots\R_{n}}
\Bigg],
\end{align}
\end{widetext}
where V$_i$ (R$_i$) indicates that the $i^\text{th}$ gluon is virtual (real), and the shorthand notation $\W_{1 \dots n}^{\{\R_j, \R_k, \R_n\}}$, for example, signifies that only the $j^\text{th}$, $k^\text{th}$, and $n^\text{th}$ gluons are real (with all other gluons being virtual). The eikonal amplitude squared in the last line, $\W_{1 \dots n}^{\R_{1}\R_{2} \dots\R_{n}}$ corresponds to the configuration where {\it all} gluons are real.
It is noteworthy that the perturbative distribution formula for the case of no clustering (such as in anti-\kt or other cone-like jet algorithms) can be readily obtained from the above expression by setting all clustering condition terms, $\O_{ij}$, to zero (and subsequently $\Ob_{ij}$ to one). Consequently, we obtain:
\begin{widetext}
\begin{align}\label{eq:nloop-NoClus-Final}
\sum_\X \Uh_n \W_{1 \dots n}^{\X} &= -\prod_{i=1}^{n} \Theta_i^v \, \Theta_n^\inn \Bigg[
\W_{1 \dots n}^{\{\R_n\}}
+ \sum_{j=1}^{n-1} \Theta_j^\out\, \W_{1 \dots n}^{ \{\R_j,\R_n\}}
+ \sum_{1\leq j<k}^{n-1} \Theta_{j}^\out \Theta_{k}^\out\,\W_{1 \dots n}^{\{\R_{j},\R_{k},\R_n\}}
\notag\\&
+ \sum_{1\leq j<k<\ell}^{n-1} \Theta_j^\out \Theta_k^\out \Theta_\ell^\out\, \W_{1 \dots n}^{\{\R_j,\R_k,\R_\ell,\R_n\}}
+  \sum_{1\leq j<k<\ell<m}^{n-1} \Theta_j^\out \Theta_k^\out \Theta_\ell^\out \Theta_m^\out\, \W_{1 \dots n}^{\{\R_j,\R_k,\R_\ell,\R_m,\R_n\}}
\notag\\&
+ \cdots
+ \Theta_{1}^\out\Theta_{2}^\out \cdots \Theta_{n-1}^\out \, \W_{1 \dots n}^{\{\R_{1},\R_{2}, \dots,\R_{n}\}}
\Bigg].
\end{align}
\end{widetext}
It can be straightforwardly verified that, up to four loops, the formula presented above for the anti-\kt (or no clustering) case reduces to the results reported in our earlier work \cite{Khelifa-Kerfa:2015mma}. Indeed, the all-orders structure of the anti-\kt clustering effect had already been evident from the findings in \cite{Khelifa-Kerfa:2015mma}. Consequently, the reduction of our \kt clustering formula \eqref{eq:nloop-Clus-Final} to the anti-\kt result serves as a robust test of its validity.

In subsequent studies \cite{Khelifa-Kerfa:2024hwx, Khelifa-Kerfa:2024gyv}\footnote{These papers were uploaded to the arXiv repository after the submission of this paper but prior to its revised version.}, we further tested Eq.~\eqref{eq:nloop-Clus-Final} by employing it to compute the distributions of specific non-global observables. These computations reaffirmed previously established results. Additionally, we compared these findings with outputs from commonly used Monte Carlo programs, observing excellent agreement.

The structure of Eq.\eqref{eq:nloop-Clus-Final} (and subsequently Eq.\eqref{eq:nloop-NoClus-Final}) underscores the {\it exponentiation} of CLs and NGLs as previously reported in \cite{Khelifa-Kerfa:2024hwx, Khelifa-Kerfa:2024gyv}, and earlier for NGLs in the anti-\kt case in \cite{Khelifa-Kerfa:2015mma}. Furthermore, the consistency of the anti-\kt formula \eqref{eq:nloop-NoClus-Final} with the Banfi-Marchesini-Smye (BMS) equation \cite{Banfi:2002hw} has been demonstrated in prior works, including Refs.~\cite{Khelifa-Kerfa:2015mma, Benslama:2020wib}.

It is noteworthy that this paper is dedicated exclusively to the structure of QCD scattering amplitudes in the eikonal approximation and does not address specific observables. For properties concerning the distribution of primary and correlated emissions relevant to typical non-global observables, we refer readers to the previously mentioned references \cite{Khelifa-Kerfa:2015mma, Benslama:2020wib, Benslama:2023gys, Khelifa-Kerfa:2024hwx, Khelifa-Kerfa:2024gyv}.

\section{Conclusion}
\label{sec:Conclusion}

In this work, we have, for the first time in the literature, derived the full analytical structure of the perturbative distribution of a generic non-global QCD observable in the context of final-state partons clustered using the (longitudinally invariant) \kt jet algorithm. This derivation is accomplished within the eikonal approximation framework, employing strong-energy ordering of the final-state partons. These approximations are sufficient to achieve single-logarithm accuracy. The resulting formulae encompass the complete color dependence and are applicable to a broad range of high-energy processes, including both leptonic and hadronic collisions.

We have explicitly derived expressions for this distribution up to four-loops through a brute force approach. This has allowed us to analyze their properties in detail and to identify a symmetric pattern that emerges consistently at two-, three-, and four-loop orders. This observed symmetry facilitates the formulation of the full distribution at the n$^\text{th}$ loop order. As a byproduct, we have determined the no-clustering case distribution by simply  switching off the clustering condition terms. Clustering effects have traditionally been treated numerically in the literature, often implemented in Monte Carlo simulations due to the intrinsic complexity of the phase space accompanying them. The analytical fixed-order structure presented here will hopefully pave the way for a deeper understanding of the mechanisms of jet clustering and its impact on various QCD observable cross-sections.

Our results corroborate previous calculations for both primary and secondary emission contributions. Unlike earlier works, our formulae are applicable to any non-global observable, encompassing the complete distribution and incorporating terms that account for CLs and NGLs. While CLs have been determined up to four-loops and NGLs up to three-loops for few specific QCD observables, our expressions can extend these calculations to higher loop orders and a wider array of observables. Some of these calculations will be published in the near future. A natural extension of this work will be to explore analogous structures for other commonly used jet algorithms, including the Cambridge/Aachen and SISCone \cite{Salam:2007xv} jet algorithms.

\begin{acknowledgments}
I would like to thank Y. Delenda for useful discussions and for reviewing the manuscript.
\end{acknowledgments}

%
\bibliography{Refs}

\begin{thebibliography}{45}%
\makeatletter
\providecommand \@ifxundefined [1]{%
 \@ifx{#1\undefined}
}%
\providecommand \@ifnum [1]{%
 \ifnum #1\expandafter \@firstoftwo
 \else \expandafter \@secondoftwo
 \fi
}%
\providecommand \@ifx [1]{%
 \ifx #1\expandafter \@firstoftwo
 \else \expandafter \@secondoftwo
 \fi
}%
\providecommand \natexlab [1]{#1}%
\providecommand \enquote  [1]{``#1''}%
\providecommand \bibnamefont  [1]{#1}%
\providecommand \bibfnamefont [1]{#1}%
\providecommand \citenamefont [1]{#1}%
\providecommand \href@noop [0]{\@secondoftwo}%
\providecommand \href [0]{\begingroup \@sanitize@url \@href}%
\providecommand \@href[1]{\@@startlink{#1}\@@href}%
\providecommand \@@href[1]{\endgroup#1\@@endlink}%
\providecommand \@sanitize@url [0]{\catcode `\\12\catcode `\$12\catcode
  `\&12\catcode `\#12\catcode `\^12\catcode `\_12\catcode `\%12\relax}%
\providecommand \@@startlink[1]{}%
\providecommand \@@endlink[0]{}%
\providecommand \url  [0]{\begingroup\@sanitize@url \@url }%
\providecommand \@url [1]{\endgroup\@href {#1}{\urlprefix }}%
\providecommand \urlprefix  [0]{URL }%
\providecommand \Eprint [0]{\href }%
\providecommand \doibase [0]{https://doi.org/}%
\providecommand \selectlanguage [0]{\@gobble}%
\providecommand \bibinfo  [0]{\@secondoftwo}%
\providecommand \bibfield  [0]{\@secondoftwo}%
\providecommand \translation [1]{[#1]}%
\providecommand \BibitemOpen [0]{}%
\providecommand \bibitemStop [0]{}%
\providecommand \bibitemNoStop [0]{.\EOS\space}%
\providecommand \EOS [0]{\spacefactor3000\relax}%
\providecommand \BibitemShut  [1]{\csname bibitem#1\endcsname}%
\let\auto@bib@innerbib\@empty
\bibitem [{\citenamefont {Catani}\ \emph {et~al.}(1993)\citenamefont {Catani},
  \citenamefont {Dokshitzer}, \citenamefont {Seymour},\ and\ \citenamefont
  {Webber}}]{Catani:1993hr}%
  \BibitemOpen
  \bibfield  {author} {\bibinfo {author} {\bibfnamefont {S.}~\bibnamefont
  {Catani}}, \bibinfo {author} {\bibfnamefont {Y.~L.}\ \bibnamefont
  {Dokshitzer}}, \bibinfo {author} {\bibfnamefont {M.~H.}\ \bibnamefont
  {Seymour}},\ and\ \bibinfo {author} {\bibfnamefont {B.~R.}\ \bibnamefont
  {Webber}},\ }\href {https://doi.org/10.1016/0550-3213(93)90166-M} {\bibfield
  {journal} {\bibinfo  {journal} {Nucl. Phys. B}\ }\textbf {\bibinfo {volume}
  {406}},\ \bibinfo {pages} {187} (\bibinfo {year} {1993})}\BibitemShut
  {NoStop}%
\bibitem [{\citenamefont {Ellis}\ and\ \citenamefont
  {Soper}(1993)}]{Ellis:1993tq}%
  \BibitemOpen
  \bibfield  {author} {\bibinfo {author} {\bibfnamefont {S.~D.}\ \bibnamefont
  {Ellis}}\ and\ \bibinfo {author} {\bibfnamefont {D.~E.}\ \bibnamefont
  {Soper}},\ }\href {https://doi.org/10.1103/PhysRevD.48.3160} {\bibfield
  {journal} {\bibinfo  {journal} {Phys. Rev.}\ }\textbf {\bibinfo {volume}
  {D48}},\ \bibinfo {pages} {3160} (\bibinfo {year} {1993})},\ \Eprint
  {https://arxiv.org/abs/hep-ph/9305266} {arXiv:hep-ph/9305266} \BibitemShut
  {NoStop}%
\bibitem [{\citenamefont {Cacciari}\ \emph {et~al.}(2012)\citenamefont
  {Cacciari}, \citenamefont {Salam},\ and\ \citenamefont
  {Soyez}}]{Cacciari:2011ma}%
  \BibitemOpen
  \bibfield  {author} {\bibinfo {author} {\bibfnamefont {M.}~\bibnamefont
  {Cacciari}}, \bibinfo {author} {\bibfnamefont {G.~P.}\ \bibnamefont
  {Salam}},\ and\ \bibinfo {author} {\bibfnamefont {G.}~\bibnamefont {Soyez}},\
  }\href {https://doi.org/10.1140/epjc/s10052-012-1896-2} {\bibfield  {journal}
  {\bibinfo  {journal} {Eur. Phys. J. C}\ }\textbf {\bibinfo {volume} {72}},\
  \bibinfo {pages} {1896} (\bibinfo {year} {2012})},\ \Eprint
  {https://arxiv.org/abs/1111.6097} {arXiv:1111.6097 [hep-ph]} \BibitemShut
  {NoStop}%
\bibitem [{\citenamefont {Cacciari}\ \emph {et~al.}(2008)\citenamefont
  {Cacciari}, \citenamefont {Salam},\ and\ \citenamefont
  {Soyez}}]{Cacciari:2008gp}%
  \BibitemOpen
  \bibfield  {author} {\bibinfo {author} {\bibfnamefont {M.}~\bibnamefont
  {Cacciari}}, \bibinfo {author} {\bibfnamefont {G.~P.}\ \bibnamefont
  {Salam}},\ and\ \bibinfo {author} {\bibfnamefont {G.}~\bibnamefont {Soyez}},\
  }\href {https://doi.org/10.1088/1126-6708/2008/04/063} {\bibfield  {journal}
  {\bibinfo  {journal} {JHEP}\ }\textbf {\bibinfo {volume} {04}},\ \bibinfo
  {pages} {063}},\ \Eprint {https://arxiv.org/abs/0802.1189} {arXiv:0802.1189
  [hep-ph]} \BibitemShut {NoStop}%
\bibitem [{\citenamefont {Dasgupta}\ and\ \citenamefont
  {Salam}(2001)}]{Dasgupta:2001sh}%
  \BibitemOpen
  \bibfield  {author} {\bibinfo {author} {\bibfnamefont {M.}~\bibnamefont
  {Dasgupta}}\ and\ \bibinfo {author} {\bibfnamefont {G.~P.}\ \bibnamefont
  {Salam}},\ }\href {https://doi.org/10.1016/S0370-2693(01)00725-0} {\bibfield
  {journal} {\bibinfo  {journal} {Phys. Lett. B}\ }\textbf {\bibinfo {volume}
  {512}},\ \bibinfo {pages} {323} (\bibinfo {year} {2001})},\ \Eprint
  {https://arxiv.org/abs/hep-ph/0104277} {arXiv:hep-ph/0104277} \BibitemShut
  {NoStop}%
\bibitem [{\citenamefont {Dasgupta}\ and\ \citenamefont
  {Salam}(2002{\natexlab{a}})}]{Dasgupta:2002bw}%
  \BibitemOpen
  \bibfield  {author} {\bibinfo {author} {\bibfnamefont {M.}~\bibnamefont
  {Dasgupta}}\ and\ \bibinfo {author} {\bibfnamefont {G.~P.}\ \bibnamefont
  {Salam}},\ }\href {https://doi.org/10.1088/1126-6708/2002/03/017} {\bibfield
  {journal} {\bibinfo  {journal} {JHEP}\ }\textbf {\bibinfo {volume} {03}},\
  \bibinfo {pages} {017}},\ \Eprint {https://arxiv.org/abs/hep-ph/0203009}
  {arXiv:hep-ph/0203009} \BibitemShut {NoStop}%
\bibitem [{\citenamefont {Appleby}\ and\ \citenamefont
  {Seymour}(2002)}]{Appleby:2002ke}%
  \BibitemOpen
  \bibfield  {author} {\bibinfo {author} {\bibfnamefont {R.~B.}\ \bibnamefont
  {Appleby}}\ and\ \bibinfo {author} {\bibfnamefont {M.~H.}\ \bibnamefont
  {Seymour}},\ }\href {https://doi.org/10.1088/1126-6708/2002/12/063}
  {\bibfield  {journal} {\bibinfo  {journal} {JHEP}\ }\textbf {\bibinfo
  {volume} {12}},\ \bibinfo {pages} {063}},\ \Eprint
  {https://arxiv.org/abs/hep-ph/0211426} {arXiv:hep-ph/0211426} \BibitemShut
  {NoStop}%
\bibitem [{\citenamefont {Banfi}\ and\ \citenamefont
  {Dasgupta}(2005)}]{Banfi:2005gj}%
  \BibitemOpen
  \bibfield  {author} {\bibinfo {author} {\bibfnamefont {A.}~\bibnamefont
  {Banfi}}\ and\ \bibinfo {author} {\bibfnamefont {M.}~\bibnamefont
  {Dasgupta}},\ }\href {https://doi.org/10.1016/j.physletb.2005.08.125}
  {\bibfield  {journal} {\bibinfo  {journal} {Phys. Lett.}\ }\textbf {\bibinfo
  {volume} {B628}},\ \bibinfo {pages} {49} (\bibinfo {year} {2005})},\ \Eprint
  {https://arxiv.org/abs/hep-ph/0508159} {arXiv:hep-ph/0508159} \BibitemShut
  {NoStop}%
\bibitem [{\citenamefont {Kelley}\ \emph
  {et~al.}(2012{\natexlab{a}})\citenamefont {Kelley}, \citenamefont {Walsh},\
  and\ \citenamefont {Zuberi}}]{Kelley:2012kj}%
  \BibitemOpen
  \bibfield  {author} {\bibinfo {author} {\bibfnamefont {R.}~\bibnamefont
  {Kelley}}, \bibinfo {author} {\bibfnamefont {J.~R.}\ \bibnamefont {Walsh}},\
  and\ \bibinfo {author} {\bibfnamefont {S.}~\bibnamefont {Zuberi}},\ }\href
  {https://doi.org/10.1007/JHEP09(2012)117} {\bibfield  {journal} {\bibinfo
  {journal} {JHEP}\ }\textbf {\bibinfo {volume} {09}},\ \bibinfo {pages}
  {117}},\ \Eprint {https://arxiv.org/abs/1202.2361} {arXiv:1202.2361 [hep-ph]}
  \BibitemShut {NoStop}%
\bibitem [{\citenamefont {Appleby}\ and\ \citenamefont
  {Seymour}(2003)}]{Appleby:2003sj}%
  \BibitemOpen
  \bibfield  {author} {\bibinfo {author} {\bibfnamefont {R.~B.}\ \bibnamefont
  {Appleby}}\ and\ \bibinfo {author} {\bibfnamefont {M.~H.}\ \bibnamefont
  {Seymour}},\ }\href {https://doi.org/10.1088/1126-6708/2003/09/056}
  {\bibfield  {journal} {\bibinfo  {journal} {JHEP}\ }\textbf {\bibinfo
  {volume} {09}},\ \bibinfo {pages} {056}},\ \Eprint
  {https://arxiv.org/abs/hep-ph/0308086} {arXiv:hep-ph/0308086} \BibitemShut
  {NoStop}%
\bibitem [{\citenamefont {Delenda}\ \emph {et~al.}(2006)\citenamefont
  {Delenda}, \citenamefont {Appleby}, \citenamefont {Dasgupta},\ and\
  \citenamefont {Banfi}}]{Delenda:2006nf}%
  \BibitemOpen
  \bibfield  {author} {\bibinfo {author} {\bibfnamefont {Y.}~\bibnamefont
  {Delenda}}, \bibinfo {author} {\bibfnamefont {R.}~\bibnamefont {Appleby}},
  \bibinfo {author} {\bibfnamefont {M.}~\bibnamefont {Dasgupta}},\ and\
  \bibinfo {author} {\bibfnamefont {A.}~\bibnamefont {Banfi}},\ }\href
  {https://doi.org/10.1088/1126-6708/2006/12/044} {\bibfield  {journal}
  {\bibinfo  {journal} {JHEP}\ }\textbf {\bibinfo {volume} {0612}},\ \bibinfo
  {pages} {044}},\ \Eprint {https://arxiv.org/abs/hep-ph/0610242}
  {arXiv:hep-ph/0610242 [hep-ph]} \BibitemShut {NoStop}%
\bibitem [{\citenamefont {Banfi}\ \emph {et~al.}(2010)\citenamefont {Banfi},
  \citenamefont {Dasgupta}, \citenamefont {Khelifa-Kerfa},\ and\ \citenamefont
  {Marzani}}]{Banfi:2010pa}%
  \BibitemOpen
  \bibfield  {author} {\bibinfo {author} {\bibfnamefont {A.}~\bibnamefont
  {Banfi}}, \bibinfo {author} {\bibfnamefont {M.}~\bibnamefont {Dasgupta}},
  \bibinfo {author} {\bibfnamefont {K.}~\bibnamefont {Khelifa-Kerfa}},\ and\
  \bibinfo {author} {\bibfnamefont {S.}~\bibnamefont {Marzani}},\ }\href
  {https://doi.org/10.1007/JHEP08(2010)064} {\bibfield  {journal} {\bibinfo
  {journal} {JHEP}\ }\textbf {\bibinfo {volume} {08}},\ \bibinfo {pages}
  {064}},\ \Eprint {https://arxiv.org/abs/1004.3483} {arXiv:1004.3483 [hep-ph]}
  \BibitemShut {NoStop}%
\bibitem [{\citenamefont {Khelifa-Kerfa}(2012)}]{Khelifa-Kerfa:2011quw}%
  \BibitemOpen
  \bibfield  {author} {\bibinfo {author} {\bibfnamefont {K.}~\bibnamefont
  {Khelifa-Kerfa}},\ }\href {https://doi.org/10.1007/JHEP02(2012)072}
  {\bibfield  {journal} {\bibinfo  {journal} {JHEP}\ }\textbf {\bibinfo
  {volume} {02}},\ \bibinfo {pages} {072}},\ \Eprint
  {https://arxiv.org/abs/1111.2016} {arXiv:1111.2016 [hep-ph]} \BibitemShut
  {NoStop}%
\bibitem [{\citenamefont {Kelley}\ \emph
  {et~al.}(2012{\natexlab{b}})\citenamefont {Kelley}, \citenamefont {Walsh},\
  and\ \citenamefont {Zuberi}}]{Kelley:2012zs}%
  \BibitemOpen
  \bibfield  {author} {\bibinfo {author} {\bibfnamefont {R.}~\bibnamefont
  {Kelley}}, \bibinfo {author} {\bibfnamefont {J.~R.}\ \bibnamefont {Walsh}},\
  and\ \bibinfo {author} {\bibfnamefont {S.}~\bibnamefont {Zuberi}},\
  }\href@noop {} {\  (\bibinfo {year} {2012}{\natexlab{b}})},\ \Eprint
  {https://arxiv.org/abs/1203.2923} {arXiv:1203.2923 [hep-ph]} \BibitemShut
  {NoStop}%
\bibitem [{\citenamefont {Delenda}\ and\ \citenamefont
  {Khelifa-Kerfa}(2012)}]{Delenda:2012mm}%
  \BibitemOpen
  \bibfield  {author} {\bibinfo {author} {\bibfnamefont {Y.}~\bibnamefont
  {Delenda}}\ and\ \bibinfo {author} {\bibfnamefont {K.}~\bibnamefont
  {Khelifa-Kerfa}},\ }\href {https://doi.org/10.1007/JHEP09(2012)109}
  {\bibfield  {journal} {\bibinfo  {journal} {JHEP}\ }\textbf {\bibinfo
  {volume} {09}},\ \bibinfo {pages} {109}},\ \Eprint
  {https://arxiv.org/abs/1207.4528} {arXiv:1207.4528 [hep-ph]} \BibitemShut
  {NoStop}%
\bibitem [{\citenamefont {Kerfa}(2012)}]{Kerfa:2012yae}%
  \BibitemOpen
  \bibfield  {author} {\bibinfo {author} {\bibfnamefont {K.~K.}\ \bibnamefont
  {Kerfa}},\ }\emph {\bibinfo {title} {{QCD resummation for high-$p_T$ jet
  shapes at hadron colliders}}},\ \href@noop {} {Ph.D. thesis},\ \bibinfo
  {school} {Manchester U.} (\bibinfo {year} {2012}),\ \Eprint
  {https://arxiv.org/abs/2111.10671} {arXiv:2111.10671 [hep-ph]} \BibitemShut
  {NoStop}%
\bibitem [{\citenamefont {Benslama}\ \emph {et~al.}(2023)\citenamefont
  {Benslama}, \citenamefont {Delenda},\ and\ \citenamefont
  {Khelifa-Kerfa}}]{Benslama:2023gys}%
  \BibitemOpen
  \bibfield  {author} {\bibinfo {author} {\bibfnamefont {H.}~\bibnamefont
  {Benslama}}, \bibinfo {author} {\bibfnamefont {Y.}~\bibnamefont {Delenda}},\
  and\ \bibinfo {author} {\bibfnamefont {K.}~\bibnamefont {Khelifa-Kerfa}},\
  }\href {https://doi.org/10.1016/j.physletb.2023.137903} {\bibfield  {journal}
  {\bibinfo  {journal} {Phys. Lett. B}\ }\textbf {\bibinfo {volume} {840}},\
  \bibinfo {pages} {137903} (\bibinfo {year} {2023})},\ \Eprint
  {https://arxiv.org/abs/2301.00860} {arXiv:2301.00860 [hep-ph]} \BibitemShut
  {NoStop}%
\bibitem [{\citenamefont {Weigert}(2004)}]{Weigert:2003mm}%
  \BibitemOpen
  \bibfield  {author} {\bibinfo {author} {\bibfnamefont {H.}~\bibnamefont
  {Weigert}},\ }\href {https://doi.org/10.1016/j.nuclphysb.2004.03.002}
  {\bibfield  {journal} {\bibinfo  {journal} {Nucl. Phys. B}\ }\textbf
  {\bibinfo {volume} {685}},\ \bibinfo {pages} {321} (\bibinfo {year}
  {2004})},\ \Eprint {https://arxiv.org/abs/hep-ph/0312050}
  {arXiv:hep-ph/0312050} \BibitemShut {NoStop}%
\bibitem [{\citenamefont {Hatta}\ and\ \citenamefont
  {Ueda}(2013)}]{Hatta:2013iba}%
  \BibitemOpen
  \bibfield  {author} {\bibinfo {author} {\bibfnamefont {Y.}~\bibnamefont
  {Hatta}}\ and\ \bibinfo {author} {\bibfnamefont {T.}~\bibnamefont {Ueda}},\
  }\href {https://doi.org/10.1016/j.nuclphysb.2013.06.021} {\bibfield
  {journal} {\bibinfo  {journal} {Nucl. Phys. B}\ }\textbf {\bibinfo {volume}
  {874}},\ \bibinfo {pages} {808} (\bibinfo {year} {2013})},\ \Eprint
  {https://arxiv.org/abs/1304.6930} {arXiv:1304.6930 [hep-ph]} \BibitemShut
  {NoStop}%
\bibitem [{\citenamefont {Hagiwara}\ \emph {et~al.}(2016)\citenamefont
  {Hagiwara}, \citenamefont {Hatta},\ and\ \citenamefont
  {Ueda}}]{Hagiwara:2015bia}%
  \BibitemOpen
  \bibfield  {author} {\bibinfo {author} {\bibfnamefont {Y.}~\bibnamefont
  {Hagiwara}}, \bibinfo {author} {\bibfnamefont {Y.}~\bibnamefont {Hatta}},\
  and\ \bibinfo {author} {\bibfnamefont {T.}~\bibnamefont {Ueda}},\ }\href
  {https://doi.org/10.1016/j.physletb.2016.03.028} {\bibfield  {journal}
  {\bibinfo  {journal} {Phys. Lett. B}\ }\textbf {\bibinfo {volume} {756}},\
  \bibinfo {pages} {254} (\bibinfo {year} {2016})},\ \Eprint
  {https://arxiv.org/abs/1507.07641} {arXiv:1507.07641 [hep-ph]} \BibitemShut
  {NoStop}%
\bibitem [{\citenamefont {Hatta}\ and\ \citenamefont
  {Ueda}(2021)}]{Hatta:2020wre}%
  \BibitemOpen
  \bibfield  {author} {\bibinfo {author} {\bibfnamefont {Y.}~\bibnamefont
  {Hatta}}\ and\ \bibinfo {author} {\bibfnamefont {T.}~\bibnamefont {Ueda}},\
  }\href {https://doi.org/10.1016/j.nuclphysb.2020.115273} {\bibfield
  {journal} {\bibinfo  {journal} {Nucl. Phys. B}\ }\textbf {\bibinfo {volume}
  {962}},\ \bibinfo {pages} {115273} (\bibinfo {year} {2021})},\ \Eprint
  {https://arxiv.org/abs/2011.04154} {arXiv:2011.04154 [hep-ph]} \BibitemShut
  {NoStop}%
\bibitem [{\citenamefont {Pl\"atzer}(2014)}]{Platzer:2013fha}%
  \BibitemOpen
  \bibfield  {author} {\bibinfo {author} {\bibfnamefont {S.}~\bibnamefont
  {Pl\"atzer}},\ }\href {https://doi.org/10.1140/epjc/s10052-014-2907-2}
  {\bibfield  {journal} {\bibinfo  {journal} {Eur. Phys. J. C}\ }\textbf
  {\bibinfo {volume} {74}},\ \bibinfo {pages} {2907} (\bibinfo {year}
  {2014})},\ \Eprint {https://arxiv.org/abs/1312.2448} {arXiv:1312.2448
  [hep-ph]} \BibitemShut {NoStop}%
\bibitem [{\citenamefont {\'Angeles~Mart\'\i{}nez}\ \emph
  {et~al.}(2018)\citenamefont {\'Angeles~Mart\'\i{}nez}, \citenamefont
  {De~Angelis}, \citenamefont {Forshaw}, \citenamefont {Pl\"atzer},\ and\
  \citenamefont {Seymour}}]{AngelesMartinez:2018cfz}%
  \BibitemOpen
  \bibfield  {author} {\bibinfo {author} {\bibfnamefont {R.}~\bibnamefont
  {\'Angeles~Mart\'\i{}nez}}, \bibinfo {author} {\bibfnamefont
  {M.}~\bibnamefont {De~Angelis}}, \bibinfo {author} {\bibfnamefont {J.~R.}\
  \bibnamefont {Forshaw}}, \bibinfo {author} {\bibfnamefont {S.}~\bibnamefont
  {Pl\"atzer}},\ and\ \bibinfo {author} {\bibfnamefont {M.~H.}\ \bibnamefont
  {Seymour}},\ }\href {https://doi.org/10.1007/JHEP05(2018)044} {\bibfield
  {journal} {\bibinfo  {journal} {JHEP}\ }\textbf {\bibinfo {volume} {05}},\
  \bibinfo {pages} {044}},\ \Eprint {https://arxiv.org/abs/1802.08531}
  {arXiv:1802.08531 [hep-ph]} \BibitemShut {NoStop}%
\bibitem [{\citenamefont {Forshaw}\ \emph {et~al.}(2019)\citenamefont
  {Forshaw}, \citenamefont {Holguin},\ and\ \citenamefont
  {Pl\"atzer}}]{Forshaw:2019ver}%
  \BibitemOpen
  \bibfield  {author} {\bibinfo {author} {\bibfnamefont {J.~R.}\ \bibnamefont
  {Forshaw}}, \bibinfo {author} {\bibfnamefont {J.}~\bibnamefont {Holguin}},\
  and\ \bibinfo {author} {\bibfnamefont {S.}~\bibnamefont {Pl\"atzer}},\ }\href
  {https://doi.org/10.1007/JHEP08(2019)145} {\bibfield  {journal} {\bibinfo
  {journal} {JHEP}\ }\textbf {\bibinfo {volume} {08}},\ \bibinfo {pages}
  {145}},\ \Eprint {https://arxiv.org/abs/1905.08686} {arXiv:1905.08686
  [hep-ph]} \BibitemShut {NoStop}%
\bibitem [{\citenamefont {Becher}\ and\ \citenamefont
  {Haag}(2024)}]{Becher:2023znt}%
  \BibitemOpen
  \bibfield  {author} {\bibinfo {author} {\bibfnamefont {T.}~\bibnamefont
  {Becher}}\ and\ \bibinfo {author} {\bibfnamefont {J.}~\bibnamefont {Haag}},\
  }\href {https://doi.org/10.1007/JHEP01(2024)155} {\bibfield  {journal}
  {\bibinfo  {journal} {JHEP}\ }\textbf {\bibinfo {volume} {01}},\ \bibinfo
  {pages} {155}},\ \Eprint {https://arxiv.org/abs/2309.17355} {arXiv:2309.17355
  [hep-ph]} \BibitemShut {NoStop}%
\bibitem [{\citenamefont {Dasgupta}\ and\ \citenamefont
  {Salam}(2002{\natexlab{b}})}]{Dasgupta:2002dc}%
  \BibitemOpen
  \bibfield  {author} {\bibinfo {author} {\bibfnamefont {M.}~\bibnamefont
  {Dasgupta}}\ and\ \bibinfo {author} {\bibfnamefont {G.~P.}\ \bibnamefont
  {Salam}},\ }\href {https://doi.org/10.1088/1126-6708/2002/08/032} {\bibfield
  {journal} {\bibinfo  {journal} {JHEP}\ }\textbf {\bibinfo {volume} {08}},\
  \bibinfo {pages} {032}},\ \Eprint {https://arxiv.org/abs/hep-ph/0208073}
  {arXiv:hep-ph/0208073} \BibitemShut {NoStop}%
\bibitem [{\citenamefont {Ziani}\ \emph {et~al.}(2021)\citenamefont {Ziani},
  \citenamefont {Khelifa-Kerfa},\ and\ \citenamefont
  {Delenda}}]{Ziani:2021dxr}%
  \BibitemOpen
  \bibfield  {author} {\bibinfo {author} {\bibfnamefont {N.}~\bibnamefont
  {Ziani}}, \bibinfo {author} {\bibfnamefont {K.}~\bibnamefont
  {Khelifa-Kerfa}},\ and\ \bibinfo {author} {\bibfnamefont {Y.}~\bibnamefont
  {Delenda}},\ }\href {https://doi.org/10.1140/epjc/s10052-021-09379-z}
  {\bibfield  {journal} {\bibinfo  {journal} {Eur. Phys. J. C}\ }\textbf
  {\bibinfo {volume} {81}},\ \bibinfo {pages} {570} (\bibinfo {year} {2021})},\
  \Eprint {https://arxiv.org/abs/2104.11060} {arXiv:2104.11060 [hep-ph]}
  \BibitemShut {NoStop}%
\bibitem [{\citenamefont {Bouaziz}\ \emph {et~al.}(2022)\citenamefont
  {Bouaziz}, \citenamefont {Delenda},\ and\ \citenamefont
  {Khelifa-Kerfa}}]{Bouaziz:2022tik}%
  \BibitemOpen
  \bibfield  {author} {\bibinfo {author} {\bibfnamefont {H.}~\bibnamefont
  {Bouaziz}}, \bibinfo {author} {\bibfnamefont {Y.}~\bibnamefont {Delenda}},\
  and\ \bibinfo {author} {\bibfnamefont {K.}~\bibnamefont {Khelifa-Kerfa}},\
  }\href {https://doi.org/10.1007/JHEP10(2022)006} {\bibfield  {journal}
  {\bibinfo  {journal} {JHEP}\ }\textbf {\bibinfo {volume} {10}},\ \bibinfo
  {pages} {006}},\ \Eprint {https://arxiv.org/abs/2207.10147} {arXiv:2207.10147
  [hep-ph]} \BibitemShut {NoStop}%
\bibitem [{\citenamefont {Delenda}\ and\ \citenamefont
  {Khelifa-Kerfa}(2016)}]{Delenda:2015tbo}%
  \BibitemOpen
  \bibfield  {author} {\bibinfo {author} {\bibfnamefont {Y.}~\bibnamefont
  {Delenda}}\ and\ \bibinfo {author} {\bibfnamefont {K.}~\bibnamefont
  {Khelifa-Kerfa}},\ }\href {https://doi.org/10.1103/PhysRevD.93.054027}
  {\bibfield  {journal} {\bibinfo  {journal} {Phys. Rev. D}\ }\textbf {\bibinfo
  {volume} {93}},\ \bibinfo {pages} {054027} (\bibinfo {year} {2016})},\
  \Eprint {https://arxiv.org/abs/1512.05401} {arXiv:1512.05401 [hep-ph]}
  \BibitemShut {NoStop}%
\bibitem [{\citenamefont {Khelifa-Kerfa}\ and\ \citenamefont
  {Delenda}(2020)}]{Khelifa-Kerfa:2020nlc}%
  \BibitemOpen
  \bibfield  {author} {\bibinfo {author} {\bibfnamefont {K.}~\bibnamefont
  {Khelifa-Kerfa}}\ and\ \bibinfo {author} {\bibfnamefont {Y.}~\bibnamefont
  {Delenda}},\ }\href {https://doi.org/10.1016/j.physletb.2020.135768}
  {\bibfield  {journal} {\bibinfo  {journal} {Phys. Lett. B}\ }\textbf
  {\bibinfo {volume} {809}},\ \bibinfo {pages} {135768} (\bibinfo {year}
  {2020})},\ \Eprint {https://arxiv.org/abs/2006.08758} {arXiv:2006.08758
  [hep-ph]} \BibitemShut {NoStop}%
\bibitem [{\citenamefont {Schwartz}\ and\ \citenamefont
  {Zhu}(2014)}]{Schwartz:2014wha}%
  \BibitemOpen
  \bibfield  {author} {\bibinfo {author} {\bibfnamefont {M.~D.}\ \bibnamefont
  {Schwartz}}\ and\ \bibinfo {author} {\bibfnamefont {H.~X.}\ \bibnamefont
  {Zhu}},\ }\href {https://doi.org/10.1103/PhysRevD.90.065004} {\bibfield
  {journal} {\bibinfo  {journal} {Phys. Rev. D}\ }\textbf {\bibinfo {volume}
  {90}},\ \bibinfo {pages} {065004} (\bibinfo {year} {2014})},\ \Eprint
  {https://arxiv.org/abs/1403.4949} {arXiv:1403.4949 [hep-ph]} \BibitemShut
  {NoStop}%
\bibitem [{\citenamefont {Khelifa-Kerfa}\ and\ \citenamefont
  {Delenda}(2015)}]{Khelifa-Kerfa:2015mma}%
  \BibitemOpen
  \bibfield  {author} {\bibinfo {author} {\bibfnamefont {K.}~\bibnamefont
  {Khelifa-Kerfa}}\ and\ \bibinfo {author} {\bibfnamefont {Y.}~\bibnamefont
  {Delenda}},\ }\href {https://doi.org/10.1007/JHEP03(2015)094} {\bibfield
  {journal} {\bibinfo  {journal} {JHEP}\ }\textbf {\bibinfo {volume} {03}},\
  \bibinfo {pages} {094}},\ \Eprint {https://arxiv.org/abs/1501.00475}
  {arXiv:1501.00475 [hep-ph]} \BibitemShut {NoStop}%
\bibitem [{\citenamefont {Dokshitzer}\ and\ \citenamefont
  {Marchesini}(2006)}]{Dokshitzer:2005ig}%
  \BibitemOpen
  \bibfield  {author} {\bibinfo {author} {\bibfnamefont {Y.}~\bibnamefont
  {Dokshitzer}}\ and\ \bibinfo {author} {\bibfnamefont {G.}~\bibnamefont
  {Marchesini}},\ }\href {https://doi.org/10.1088/1126-6708/2006/01/007}
  {\bibfield  {journal} {\bibinfo  {journal} {JHEP}\ }\textbf {\bibinfo
  {volume} {0601}},\ \bibinfo {pages} {007}},\ \Eprint
  {https://arxiv.org/abs/hep-ph/0509078} {arXiv:hep-ph/0509078 [hep-ph]}
  \BibitemShut {NoStop}%
\bibitem [{\citenamefont {Forshaw}\ \emph {et~al.}(2006)\citenamefont
  {Forshaw}, \citenamefont {Kyrieleis},\ and\ \citenamefont
  {Seymour}}]{Forshaw:2006fk}%
  \BibitemOpen
  \bibfield  {author} {\bibinfo {author} {\bibfnamefont {J.~R.}\ \bibnamefont
  {Forshaw}}, \bibinfo {author} {\bibfnamefont {A.}~\bibnamefont {Kyrieleis}},\
  and\ \bibinfo {author} {\bibfnamefont {M.~H.}\ \bibnamefont {Seymour}},\
  }\href {https://doi.org/10.1088/1126-6708/2006/08/059} {\bibfield  {journal}
  {\bibinfo  {journal} {JHEP}\ }\textbf {\bibinfo {volume} {08}},\ \bibinfo
  {pages} {059}},\ \Eprint {https://arxiv.org/abs/hep-ph/0604094}
  {arXiv:hep-ph/0604094} \BibitemShut {NoStop}%
\bibitem [{\citenamefont {Becher}\ \emph {et~al.}(2021)\citenamefont {Becher},
  \citenamefont {Neubert},\ and\ \citenamefont {Shao}}]{Becher:2021zkk}%
  \BibitemOpen
  \bibfield  {author} {\bibinfo {author} {\bibfnamefont {T.}~\bibnamefont
  {Becher}}, \bibinfo {author} {\bibfnamefont {M.}~\bibnamefont {Neubert}},\
  and\ \bibinfo {author} {\bibfnamefont {D.~Y.}\ \bibnamefont {Shao}},\ }\href
  {https://doi.org/10.1103/PhysRevLett.127.212002} {\bibfield  {journal}
  {\bibinfo  {journal} {Phys. Rev. Lett.}\ }\textbf {\bibinfo {volume} {127}},\
  \bibinfo {pages} {212002} (\bibinfo {year} {2021})},\ \Eprint
  {https://arxiv.org/abs/2107.01212} {arXiv:2107.01212 [hep-ph]} \BibitemShut
  {NoStop}%
\bibitem [{\citenamefont {Becher}\ \emph {et~al.}(2023)\citenamefont {Becher},
  \citenamefont {Neubert}, \citenamefont {Shao},\ and\ \citenamefont
  {Stillger}}]{Becher:2023mtx}%
  \BibitemOpen
  \bibfield  {author} {\bibinfo {author} {\bibfnamefont {T.}~\bibnamefont
  {Becher}}, \bibinfo {author} {\bibfnamefont {M.}~\bibnamefont {Neubert}},
  \bibinfo {author} {\bibfnamefont {D.~Y.}\ \bibnamefont {Shao}},\ and\
  \bibinfo {author} {\bibfnamefont {M.}~\bibnamefont {Stillger}},\ }\href
  {https://doi.org/10.1007/JHEP12(2023)116} {\bibfield  {journal} {\bibinfo
  {journal} {JHEP}\ }\textbf {\bibinfo {volume} {12}},\ \bibinfo {pages}
  {116}},\ \Eprint {https://arxiv.org/abs/2307.06359} {arXiv:2307.06359
  [hep-ph]} \BibitemShut {NoStop}%
\bibitem [{\citenamefont {B\"oer}\ \emph {et~al.}(2024)\citenamefont {B\"oer},
  \citenamefont {Hager}, \citenamefont {Neubert}, \citenamefont {Stillger},\
  and\ \citenamefont {Xu}}]{Boer:2024hzh}%
  \BibitemOpen
  \bibfield  {author} {\bibinfo {author} {\bibfnamefont {P.}~\bibnamefont
  {B\"oer}}, \bibinfo {author} {\bibfnamefont {P.}~\bibnamefont {Hager}},
  \bibinfo {author} {\bibfnamefont {M.}~\bibnamefont {Neubert}}, \bibinfo
  {author} {\bibfnamefont {M.}~\bibnamefont {Stillger}},\ and\ \bibinfo
  {author} {\bibfnamefont {X.}~\bibnamefont {Xu}},\ }\href
  {https://doi.org/10.1007/JHEP08(2024)035} {\bibfield  {journal} {\bibinfo
  {journal} {JHEP}\ }\textbf {\bibinfo {volume} {08}},\ \bibinfo {pages}
  {035}},\ \Eprint {https://arxiv.org/abs/2405.05305} {arXiv:2405.05305
  [hep-ph]} \BibitemShut {NoStop}%
\bibitem [{\citenamefont {Dokshitzer}\ \emph {et~al.}(1997)\citenamefont
  {Dokshitzer}, \citenamefont {Leder}, \citenamefont {Moretti},\ and\
  \citenamefont {Webber}}]{Dokshitzer:1997in}%
  \BibitemOpen
  \bibfield  {author} {\bibinfo {author} {\bibfnamefont {Y.~L.}\ \bibnamefont
  {Dokshitzer}}, \bibinfo {author} {\bibfnamefont {G.~D.}\ \bibnamefont
  {Leder}}, \bibinfo {author} {\bibfnamefont {S.}~\bibnamefont {Moretti}},\
  and\ \bibinfo {author} {\bibfnamefont {B.~R.}\ \bibnamefont {Webber}},\
  }\href {https://doi.org/10.1088/1126-6708/1997/08/001} {\bibfield  {journal}
  {\bibinfo  {journal} {JHEP}\ }\textbf {\bibinfo {volume} {08}},\ \bibinfo
  {pages} {001}},\ \Eprint {https://arxiv.org/abs/hep-ph/9707323}
  {arXiv:hep-ph/9707323} \BibitemShut {NoStop}%
\bibitem [{\citenamefont {Wobisch}\ and\ \citenamefont
  {Wengler}(1998)}]{Wobisch:1998wt}%
  \BibitemOpen
  \bibfield  {author} {\bibinfo {author} {\bibfnamefont {M.}~\bibnamefont
  {Wobisch}}\ and\ \bibinfo {author} {\bibfnamefont {T.}~\bibnamefont
  {Wengler}},\ }in\ \href@noop {} {\emph {\bibinfo {booktitle} {{Workshop on
  Monte Carlo Generators for HERA Physics (Plenary Starting Meeting)}}}}\
  (\bibinfo {year} {1998})\ pp.\ \bibinfo {pages} {270--279},\ \Eprint
  {https://arxiv.org/abs/hep-ph/9907280} {arXiv:hep-ph/9907280} \BibitemShut
  {NoStop}%
\bibitem [{\citenamefont {Khelifa-Kerfa}(2024{\natexlab{a}})}]{CAclus}%
  \BibitemOpen
  \bibfield  {author} {\bibinfo {author} {\bibfnamefont {K.}~\bibnamefont
  {Khelifa-Kerfa}}} (\bibinfo {year} {2024}{\natexlab{a}}),\ \bibinfo {note}
  {{in progress}}\BibitemShut {NoStop}%
\bibitem [{\citenamefont
  {Khelifa-Kerfa}(2024{\natexlab{b}})}]{Khelifa-Kerfa:2024hwx}%
  \BibitemOpen
  \bibfield  {author} {\bibinfo {author} {\bibfnamefont {K.}~\bibnamefont
  {Khelifa-Kerfa}},\ }\href {https://arxiv.org/abs/2411.03956} {\bibinfo
  {title} {Dijet mass up to four-loops with(out) ${\boldsymbol k}_{\boldsymbol
  t}$ clustering}} (\bibinfo {year} {2024}{\natexlab{b}}),\ \Eprint
  {https://arxiv.org/abs/2411.03956} {arXiv:2411.03956 [hep-ph]} \BibitemShut
  {NoStop}%
\bibitem [{\citenamefont
  {Khelifa-Kerfa}(2024{\natexlab{c}})}]{Khelifa-Kerfa:2024gyv}%
  \BibitemOpen
  \bibfield  {author} {\bibinfo {author} {\bibfnamefont {K.}~\bibnamefont
  {Khelifa-Kerfa}},\ }\href {https://arxiv.org/abs/2412.03244} {\bibinfo
  {title} {Clustering logarithms up to six loops}} (\bibinfo {year}
  {2024}{\natexlab{c}}),\ \Eprint {https://arxiv.org/abs/2412.03244}
  {arXiv:2412.03244 [hep-ph]} \BibitemShut {NoStop}%
\bibitem [{\citenamefont {Banfi}\ \emph {et~al.}(2002)\citenamefont {Banfi},
  \citenamefont {Marchesini},\ and\ \citenamefont {Smye}}]{Banfi:2002hw}%
  \BibitemOpen
  \bibfield  {author} {\bibinfo {author} {\bibfnamefont {A.}~\bibnamefont
  {Banfi}}, \bibinfo {author} {\bibfnamefont {G.}~\bibnamefont {Marchesini}},\
  and\ \bibinfo {author} {\bibfnamefont {G.}~\bibnamefont {Smye}},\ }\href
  {https://doi.org/10.1088/1126-6708/2002/08/006} {\bibfield  {journal}
  {\bibinfo  {journal} {JHEP}\ }\textbf {\bibinfo {volume} {08}},\ \bibinfo
  {pages} {006}},\ \Eprint {https://arxiv.org/abs/hep-ph/0206076}
  {arXiv:hep-ph/0206076} \BibitemShut {NoStop}%
\bibitem [{\citenamefont {Benslama}\ \emph {et~al.}(2021)\citenamefont
  {Benslama}, \citenamefont {Delenda}, \citenamefont {Khelifa-Kerfa},\ and\
  \citenamefont {Ibrahim}}]{Benslama:2020wib}%
  \BibitemOpen
  \bibfield  {author} {\bibinfo {author} {\bibfnamefont {H.}~\bibnamefont
  {Benslama}}, \bibinfo {author} {\bibfnamefont {Y.}~\bibnamefont {Delenda}},
  \bibinfo {author} {\bibfnamefont {K.}~\bibnamefont {Khelifa-Kerfa}},\ and\
  \bibinfo {author} {\bibfnamefont {A.~M.}\ \bibnamefont {Ibrahim}},\ }\href
  {https://doi.org/10.1134/S1547477121010039} {\bibfield  {journal} {\bibinfo
  {journal} {Phys. Part. Nucl. Lett.}\ }\textbf {\bibinfo {volume} {18}},\
  \bibinfo {pages} {5} (\bibinfo {year} {2021})},\ \Eprint
  {https://arxiv.org/abs/2006.06738} {arXiv:2006.06738 [hep-ph]} \BibitemShut
  {NoStop}%
\bibitem [{\citenamefont {Salam}\ and\ \citenamefont
  {Soyez}(2007)}]{Salam:2007xv}%
  \BibitemOpen
  \bibfield  {author} {\bibinfo {author} {\bibfnamefont {G.~P.}\ \bibnamefont
  {Salam}}\ and\ \bibinfo {author} {\bibfnamefont {G.}~\bibnamefont {Soyez}},\
  }\href {https://doi.org/10.1088/1126-6708/2007/05/086} {\bibfield  {journal}
  {\bibinfo  {journal} {JHEP}\ }\textbf {\bibinfo {volume} {05}},\ \bibinfo
  {pages} {086}},\ \Eprint {https://arxiv.org/abs/0704.0292} {arXiv:0704.0292
  [hep-ph]} \BibitemShut {NoStop}%
\end{thebibliography}%

\end{document}